# Improved Carrier Mobility in Few-Layer MoS$_2$ Field-Effect Transistors with Ionic-Liquid Gating

Meeghage Madusanka Perera,[1] Ming-Wei Lin,[1] Hsun-Jen Chuang,[1] Bhim Prasad Chamlagain,[1] Chongyu Wang,[1] Xuebin Tan,[2] Mark Ming-Cheng Cheng,[2] David Tománek,[3] and Zhixian Zhou[1, a)]

[1]Department of Physics and Astronomy, Wayne State University, Detroit, MI 48201

[2]Department of Electrical and Computer Engineering, Wayne State University, Detroit, MI 48202

3 Physics and Astronomy Department, Michigan State University, East Lansing, MI 48824

Abstract

We report the fabrication of ionic liquid (IL) gated field-effect transistors (FETs) consisting of bilayer and few-layer MoS$_2$. Our transport measurements indicate that the electron mobility µ≈60 cm$^2$V$^{-1}$s$^{-1}$ at 250 K in ionic liquid gated devices exceeds significantly that of comparable back-gated devices. IL-FETs display a mobility increase from ≈100 cm$^2$V$^{-1}$s$^{-1}$ at 180 K to ≈220 cm$^2$V$^{-1}$s$^{-1}$ at 77 K in good agreement with the true channel mobility determined from four-terminal measurements, ambipolar behavior with a high ON/OFF ratio >10$^7$ (10$^4$) for electrons (holes), and a near ideal sub-threshold swing of ≈50 mV/dec at 250 K. We attribute the observed performance enhancement, specifically the increased carrier mobility that is limited by phonons, to the reduction of the Schottky barrier at the source and drain electrode by band bending caused by the ultrathin ionic-liquid dielectric layer.

a) Author to whom correspondence should be addressed, electronic mail: zxzhou@wayne.edu

Key words: Field-effect transistor, MoS$_2$, few-layer, electric double layer, Schottky barrier

In the quest for flexible electronics in the "post-silicon" era, graphene has attracted much attention due to unsurpassed carrier mobility and high thermal conductivity,[1-4] combined with excellent chemical and thermal stability down to the nanometer scale.[5] The major drawback is the absence of fundamental band gap, which makes semimetallic graphene unsuitable for conventional digital logic applications. Sustained efforts to engineer a band gap in graphene have either caused severe mobility degradation or require prohibitively high bias voltages.[6-9]

Molybdenum disulfide ($MoS_2$), a layered transition-metal dichalcogenide (TMD), has emerged as a viable alternative to graphene, as it combines a semiconducting gap with mechanical flexibility, chemical and thermal stability and absence of dangling bonds. The single-layer $MoS_2$ consists of a molybdenum monolayer sandwiched between two sulfur monolayers. The fundamental band gap changes from an ≈1.2 eV wide indirect gap in the bulk to a direct gap of ≈1.8 eV in single-layer $MoS_2$.[10-11] Similar to graphene, single $MoS_2$ layers can be extracted from bulk crystals by a mechanical cleavage technique due to relatively weak interlayer interaction with an important van der Waals character.[12] Besides conventional field effect transistors (FETs), the use of $MoS_2$ has been proposed for applications such as energy harvesting[13-14] and optoelectronics.[15-16] Recently, integrated circuits based on $MoS_2$ transistors have also been demonstrated, which is a significant step toward the application of $MoS_2$ in high-performance low-power nanoelectronics.[17] However, the room temperature carrier mobility in single- and few-layer $MoS_2$ FETs fabricated on $Si/SiO_2$ substrates was found to be very low, typically in the range of 0.1 -10 $cm^2V^{-1}s^{-1}$.[12, 17, 21] This mobility is not only orders of magnitude lower than that of graphene, but also substantially lower than the phonon-limited mobility in the bulk system,[12, 18-21] which is of the order of 100 $cm^2V^{-1}s^{-1}$. The interface between the $MoS_2$ channel and the $SiO_2$ gate dielectric has been considered as one of the primary factors limiting carrier mobility.[18] A substantial mobility enhancement to over 200 $cm^2V^{-1}s^{-1}$ and 500 $cm^2V^{-1}s^{-1}$ has been reported for $HfO_2$ and $Al_2O_3$ capped monolayer and multilayer $MoS_2$ FETs, respectively, which was attributed to high-κ dielectric screening of charged impurities that reduces scattering at the channel/dielectric interface.[22-23] There are also rising concerns that the mobility in these devices might have been substantially overestimated.[24]

For large band-gap semiconductors such as $MoS_2$, a significant Schottky barrier may form at the metal/semiconductor contact, yielding a high contact resistance.[10,25] Lee *et al.* showed in their study of $MoS_2$ flakes produced by liquid exfoliation that the mobility in $MoS_2$ FETs could be largely underestimated due to the Schottky barriers at the $MoS_2$/metal contacts.[26] In the presence of a substantial Schottky barrier, the extrinsic mobility is also expected to degrade with decreasing temperature due to the reduced thermionic emission current and thermally assisted tunneling current, as was recently observed by Ghatak *et al* in atomically thin $MoS_2$ FETs.[18] In agreement with recent predictions,[24] Das *et al*. has demonstrated a significant mobility enhancement by reducing the Schottky barrier height using a low work function contact metal, which further indicates that the performance of $MoS_2$ FETs can be strongly influenced by the metal/semiconductor contacts.[26]

In order to optimize the performance of $MoS_2$ FETs, it is crucial to use low resistance Ohmic contacts. There are typically two types of low resistance contacts that can be made between a semiconductor and a metal: (a) Schottky contacts with a very low barrier height and (b) highly transparent tunneling contacts. Ideally, an Ohmic contact can be formed if the Schottky barrier height is zero (or negative). Contacts with low Schottky barrier height (≈30 meV) have been achieved in multilayer $MoS_2$ by using Scandium as a low work function contact metal.[27] However, the tunability of the Schottky barrier height may be reduced by Fermi level pinning.[21] Alternatively, highly transparent tunneling contacts can be fabricated by heavily doping the semiconductor in the contact region. This approach fails for $MoS_2$, since ionized impurity doping would substantially damage the structural integrity of the atomically thin channel. As an alternative, surface doping with strongly oxidizing $NO_2$ molecules has been used to narrow the Schottky barrier thickness for hole injection and thus reduce the contact resistance of $WSe_2$ FETs.[28]

In this article, we report electrostatic doping using an ionic-liquid (IL) gate as a viable approach to achieve low resistance $MoS_2$/metal tunneling contacts. We demonstrate (i) significant improvement in the performance of few-layer $MoS_2$ FETs, and (ii) high carrier mobility in the $MoS_2$ channel that is limited by phonons. Ionic liquids are binary organic salts that can form electric double layers at the ionic-liquid/solid interface and thus act as nano-gap capacitors with extremely large capacitance. As we show in the following, the Schottky barrier can be drastically reduced in ionic-liquid-gated FETs (IL-FETs) of $MoS_2$. We observe a significant increase of the tunneling efficiency that can be attributed to strong band bending at the $MoS_2$/metal interface, provided by the thin electrical double layer with a high capacitance. As a result, our nanometer-thick $MoS_2$ IL-FETs exhibit a significantly enhanced extrinsic mobility that exceeds 60 $cm^2V^{-1}s^{-1}$ at 250 K, in contrast to µ<5 $cm^2V^{-1}s^{-1}$ measured in the Si-back-gate configuration without ionic liquid. The $MoS_2$ IL-FETs further exhibit ambipolar behavior with a high current ON/OFF ratio exceeding $10^7$ for electrons and $10^4$ for holes, and a near ideal subthreshold swing (SS) of ~50 mV/decade at 250 K. More significantly, the mobility in few-layer $MoS_2$ IL-FETs increases from ~100 $cm^2V^{-1}s^{-1}$ to ~220 $cm^2V^{-1}s^{-1}$ as the temperature decreases from 180 K to 77 K, which is in good agreement with the true channel mobility derived from our four-terminal measurements. The temperature dependence of the mobility behaves as $\mu \sim T^{-\gamma}$ with $\gamma \approx 1$, indicating that the mobility is predominantly limited by phonon scattering in this case.

**RESULTS AND DISCUSSION**

Atomically thin MoS$_2$ flakes were produced from a bulk crystal by a mechanical cleavage method and subsequently transferred onto degenerately doped silicon substrates covered with a 290 nm-thick thermal oxide layer.[12, 29] An optical microscope was used to identify thin flakes, which were further characterized by non-contact mode atomic force microscopy (AFM). In the present study, we focus on bilayer and few-layer (2-7 layers corresponding to 1.3-5 nm thickness) samples, since the yield of bilayer and few-layer flakes was found to be much higher than that of single-layer MoS$_2$. Moreover, few-layer MoS$_2$ also tends to form lower Schottky barriers (thus smaller contact resistance) than single-layer samples.[21, 27] MoS$_2$ IL-FET devices were fabricated by first patterning the source, drain and gate electrodes, consisting of 5 nm of Ti covered by 50 nm of Au, using standard electron beam lithography and electron beam deposition.[8] A small droplet of the DEME-TFSI ionic liquid (Sigma Aldrich 727679) was then carefully applied onto the devices using a micromanipulator under an optical microscope, covering the MoS$_2$ layer and the source, drain and gate electrodes.[30] The ionic liquid gate forms by self-assembly. DEME-TFSI has been chosen for its large electrochemical stability window (>3 V at room temperature).

Figure 1a shows a micrograph and figure 1b the schematic of a typical ionic-liquid-gated MoS$_2$ device. Electrical properties of the devices were measured by a Keithley 4200 semiconductor parameter analyzer in a Lakeshore Cryogenic probe station after dehydrating the ionic liquid under high vacuum (~1×10$^{-6}$ Torr) for 48 hours. This thorough removal of the remaining moisture turned out to be important to preventing the formation of chemically reactive protons and hydroxyls through the electrolysis of water.[31] Most measurements on ionic-liquid-gated devices were carried out at 250 K or below to further reduce the possibility of any chemical reactions between the ionic liquid and $MoS_2$.[32] As shown schematically in figure 1b, negative ions in the ionic liquid accumulate near the gate electrode and positive ions accumulate near the $MoS_2$ channel when a positive voltage is applied to an ionic-liquid-gate-electrode near the device channel. The scenario reverses when a negative voltage is applied to the gate. In both cases, electric double layers form at the interfaces between the ionic liquid and solid surfaces.[33] To ensure that nearly all of the gate voltage appears as potential drop across the ionic-liquid/channel interface, the surface area of the gate electrode is 1-2 orders of magnitude larger than the total area of the transport channel plus parts of the drain/source electrodes, which are immersed in the ionic liquid.[34] Downscaling of the ionic liquid gated devices can be achieved by simultaneously reducing the surface area of the gate electrode and covering a large part of the drain/source electrodes with an insulating overlayer.[35]

We have measured several ionic-liquid-gated bilayer and few-layer $MoS_2$ FETs and observed consistent results. Fig. 2 shows the transfer characteristics of two representative devices measured at a drain-source voltage of 1 V. Both the bilayer and trilayer devices exhibit ambipolar behavior, with the current ON/OFF ratio exceeding $10^7$ for electrons in both devices. The observed ON/OFF ratio for holes was $10^6$ in the bilayer and $10^4$ in the trilayer device. Ambipolar behavior has been previously observed in ion-liquid-gated thicker $MoS_2$ flakes (>10 nm) by Zhang *et. al*.[32] However, their ON/OFF ratio was less than $10^3$ for both electrons and holes, presumably due to the relatively large "OFF" state current passing through the interior of the crystal beneath the channel surface. This current ON/OFF ratio is much lower than the typical values between $10^4$ and $10^7$, which are desired for digital logic devices.[22] It is worth pointing out that our observation of hole conduction in bilayer and few-layer MoS$_2$ is rather surprising in view of the large Schottky barrier height (~1 eV) for the hole-channel.[21] Our results suggest that holes are injected into the MoS$_2$ channel primarily by thermally assisted tunneling rather than by thermionic emission.[36] The tunneling rate is increased significantly for both electrons and holes in presence of the extremely thin (~1 nm) dielectric layer formed by the ionic-liquid gate, which significantly reduces the thickness of Schottky barrier through strong band bending near the contacts at high gate voltages. Since the formation of an electrical double layer on the MoS$_2$ contacts near the edge of the metal electrodes is conformal, the thickness of the contact Schottky barrier can be reduced very effectively by the ionic liquid gate down to the electrostatic screening length in the ionic liquid (~1 nm).[37] The asymmetry between electron and hole transport can be attributed to 1) a larger Schottky barrier height for the hole channel that reduces thermally assisted tunneling, 2) a slight preference for the adsorption of positive ions on MoS$_2$, as discussed later, and 3) intrinsic *n*-doping of the transport channel. All of these effects tend to favor electron *vs.* hole transport, shifting the transfer curves toward the negative gate-voltage direction. The lower asymmetry between electron and hole transport (observed when the gate voltage was swept from positive to negative), causing a more balanced ambipolar character of the bilayer MoS$_2$ device, may be attributed to a slightly lower degree of intrinsic *n*-doping in the bilayer flake.[34] The hysteresis in the transfer characteristics could be attributed to the charge injection at the interfaces between the channel and the substrate as well as the slow motion of the ions at low temperature.[32, 38]

Low contact resistance is an important prerequisite to realize the full potential of MoS$_2$ as a channel material for FETs. Since the Schottky barrier for holes is larger than for electrons, the contact resistance in the hole channel is higher than in the electron channel. To optimize the device performance, we next focus on the electron channel only and study the impact of ionic-liquid-induced Schottky barrier thinning on its electrical characteristics. Fig. 3 shows the output characteristics of a trilayer MoS$_2$ device that was measured both with an ionic-liquid-gate and a back-gate with no ionic liquid present. As shown in Fig. 3a, the drain current in the ionic-liquid gate voltage range of $0 < V_{ILg} < 1$ V exhibits linear dependence at low drain-source voltages and starts to saturate at higher $V_{ds}$. The current saturation at high $V_{ds}$ can be attributed to the channel pinch-off of the FET. In sharp contrast to these data, the same device, when measured in the back-gate configuration without ionic liquid, exhibits strongly non-linear (upward turning) $I_{ds}$–$V_{ds}$ behavior, suggesting the presence of a significant Schottky barrier at the contacts (Fig. 3b). Furthermore, the total resistance calculated from the slope of the $I_{ds}$ - $V_{ds}$ characteristics in the low-bias region is over two orders of magnitude larger for the Si back-gate configuration ($2\times10^6$ Ω at $V_{bg}$ = 60 V, see the inset of Fig. 3b) than for the ionic-liquid-gate configuration ($1\times10^4$ Ω at $V_{ILg}$ = 1 V), providing further evidence that ionic-liquid gating significantly reduces the contact resistance by thinning the Schottky barrier. Note that linear $I_{ds}$-$V_{ds}$ dependence at small bias voltages ($V_{ds} < 0.1$ V, shown in the inset of Fig. 3b) is only a necessary, but not a sufficient condition for a low-resistance Ohmic contact. Linear current-voltage behavior may also be due to the thermally assisted tunneling current, especially at small drain-source voltages.[27]

The observed drastic reduction of the contact resistance by ionic-liquid gating opens up the possibility of investigating channel-limited device parameters in nanometer-thick MoS$_2$ devices. Fig. 4a shows the transfer characteristics from two separate measurements of the same trilayer MoS$_2$ device at T= 250 K, for $V_{ds}$ = 0.1 V and $V_{bg}$ = 0 V. The high reproducibility of the transfer curves indicate that charged ions in the ionic liquid are electrostatically accumulated at the gate/electrolyte and $MoS_2$/electrolyte interfaces without any noticeable chemical reactions. The transfer characteristics also remain essentially unchanged at different gate voltage sweeping rates. Furthermore, the subthreshold swing (SS) reaches the theoretical limit of $kT/e \ln(10) = 50$ meV/decade at T = 250 K corresponding to a gate efficiency of ~1. Such a high gate efficiency can be attributed to the large electric-double-layer capacitance of the ionic-liquid gate. The near ideal subthreshold swing also further indicates that the ionic-liquid gate creates highly transparent tunneling contacts.[39]

To extract the carrier mobility, we first estimated the ionic-liquid gate capacitance by measuring $I_{ds}$ versus $V_{ILg}$ of the same trilayer device at various fixed back-gate voltages, as shown in Fig. 4b. As the back-gate voltage is stepped up from the 0 to 30 V, the threshold voltage $V_{th}$ of the $I_{ds}$-$V_{ILg}$ curves systematically shifts in the negative $V_{ILg}$ direction, while the slope of the $I_{ds}$-$V_{ILg}$ curves remains nearly constant in the linear region. The small crossover between the $I_{ds}$-$V_{ILg}$ curve measured at $V_{bg}$ = 30 V and corresponding measurements at lower back-gate voltages may be due to the hysteretic effect. The ionic-liquid gate capacitance per unit area is estimated to be $C_{ILg}$ ~ $1.55\times10^{-6}$ Fcm$^{-2}$. This estimate is based on the observed change of the threshold voltage $\Delta V_{th}$ in response to the change of the back-gate voltage $\Delta V_{bg}$ using the relationship $C_{ILg}/C_{bg} = \Delta V_{bg}/\Delta V_{th}$. We used $C_{bg}$ = $1.2\times10^{-8}$ Fcm$^{-2}$ for the capacitance per unit area between the channel and the back gate and the value $\Delta V_{bg}/\Delta V_{th}$ = 129, determined from the linear fit shown in the inset of Fig. 4b. Using the expression $\mu = L/W \times dI_{ds}/dV_{ILg}/(C_{ILg}V_{ds})$, we estimated the low-field field-effect mobility of ~293 cm$^2$V$^{-1}$s$^{-1}$ using $L$ = 3.3 μm for the channel length and $W$ = 1.0 μm for the channel width, $dI_{ds}/dV_{tg}$ for the slope of $I_{ds}$-$V_{ILg}$ curve in the linear region at $V_{bg}$ = 0 V, and $C_{ILg}$ ~$1.55\times10^{-6}$ F/cm$^{-2}$. Note that the value $C_{ILg}$ ~ $1.55\times10^{-6}$ F/cm$^{-2}$ is about 4-5 times smaller than the $C_{ILg}$ value determined by Hall measurements ($C_{ILg,H}$ ~ $7.2\times10^{-6}$

Fcm$^{-2}$) on much thicker MoS$_2$ flakes[32]. The discrepancy may arise from the dependence of the quantum capacitance on the carrier density. The total capacitance $C_{ILg}$, consisting of the electrostatic capacitance $C_e$ of the electric double layer and the quantum capacitance $C_q$ of the MoS$_2$ channel, which are connected in series ($1/C_{ILg}= 1/C_e+1/C_q$). $C_{ILg}$, is likely dominated by $C_q$ due to the extremely large electrostatic capacitance of the DEME-TFSI ionic liquid gate,[31] which can be as high as 100 µF/cm$^2$. Since $C_q$ is a measure of the average density of states (DOS) at the Fermi level, which increases with increasing carrier density, the value of $C_q$ may also increase with the carrier density.[10] As a result, the $C_q$ is expected to be smaller in the low carrier density region near the threshold voltage than in the higher carrier density region [above $n_{2D}$~1×10$^{13}$ cm$^2$ as determined from $n_{2D} = C_{ILg,H}(V_{ILg,H} -V_{th})$], where the field-effect mobility is determined. This may lead to a possible underestimate of the total capacitance and thus an overestimate of the mobility. Using the capacitance $C_{ILg,H}$ = 7.2 µFcm$^{-2}$ determined by Hall measurement in multilayer flakes ( > 10 nm) at high carrier densities, the low limit of the field-effect mobility in our ionic-liquid-gated trilayer MoS$_2$ device is determined to be 63 cm$^2$V$^{-1}$s$^{-1}$, consistent with the Hall mobility measured in ionic-liquid-gated MoS$_2$ multiplayer flakes and with the mobility of multiplayer MoS$_2$ on SiO$_2$ measured in a four-probe configuration.[32, 40] We conclude that the actual extrinsic field-effect mobility lies likely between 63 cm$^2$V$^{-1}$s$^{-1}$ and 293 cm$^2$V$^{-1}$s$^{-1}$, which is 1-2 orders of magnitude higher than the mobility observed in typical Si-back-gated monolayer and few-layer MoS$_2$ FETs. This indicates the reported mobility ranging between 0.1–10 cm$^2$V$^{-1}$s$^{-1}$ in monolayer and few-layer MoS$_2$ FETs has been largely limited by the contact resistance. [12, 18, 22]

It is also worth noting that our ionic-liquid-gated MoS$_2$ channel is in a highly electron-doped state with a threshold voltage of $V_{th} \approx 0.5$ V at $V_{ILg} = 0$ V, which may be attributed to a higher concentration of positive than negative ions adsorbed in the vicinity of MoS$_2$. A large negative threshold voltage shift was also observed in high-κ dielectric passivated monolayer and multiplayer MoS$_2$ FETs, which could be attributed to the presence of a large amount of fixed positive charges in the dielectric layer, which have likely accumulated during the low temperature atomic layer deposition process.[22-23] Similar to the molecular ions adsorbed on the MoS$_2$ surface in ionic-liquid gated devices, these fixed charges in the thin high-κ dielectric layer could also reduce the Schottky barrier thickness, and thus contribute to the reported mobility enhancement. Also in carbon nanotube FETs, modest surface molecular doping has been shown to significantly reduce the Schottky barrier thickness, leading to a substantially enhanced tunneling current.[41]

To elucidate the transport mechanisms in the few-layer MoS$_2$ channel that was electrostatically doped by the ionic liquid, we measured the $I_{ds}$-$V_{bg}$ relationship in a different MoS$_2$ device (3.3 nm or 5 layers thick) between 77 and 180K, after the device had been quickly cooled from 250 K to 77 K at a fixed $V_{ILg}$ = 0 V. Below the freezing point of the ionic liquid (≈200 K), the carrier charge density induced by the presence of positive ions, which preferentially enriched the vicinity of MoS$_2$, remained practically constant. The carefully chosen value $V_{ILg}$ = 0 V of the ionic-liquid gate voltage allowed the creation of highly transparent tunneling contacts due to the adsorption of positive ions (implying *n*-doping), while the carrier density in the MoS$_2$ channel was kept low enough to allow an efficient reduction to zero by the back gate (see Fig. 5a). As shown in the inset of Fig. 5a, the $I_{ds}$-$V_{ds}$ characteristics are highly linear in the entire $V_{ds}$ and $V_{bg}$ range even at 77 K, indicating highly transparent contacts with thin Schottky barriers. To confirm that it is the electrostatic surface doping that is responsible for the drastic reduction of the contact resistance, we compared the device characteristics before the ionic liquid was added and after it was removed, following the completion of all electrical measurements. We observed nearly identical output characteristics in both cases (data not shown). Consequently, we may exclude the possibility of electrochemical doping or any other type of irreversible electrochemically induced degradation of the MoS$_2$ channel.

To avoid possible complications arising from the interplay between the ionic-liquid gate and the back gate, we performed $I_{ds}$-$V_{bg}$ measurements only at temperatures below the freezing temperature of the ionic liquid in order to suppress changes in the capacitance between the back-gate and the MoS$_2$ channel. In our previous study of polymer-electrolyte-gated monolayer-thick MoS$_2$ FETs, we attributed the beneficial influence of the polymer-electrolyte gate on the $I_{ds}$-$V_{bg}$ curves to a significant improvement of the carrier mobility.[42] However, the extent of mobility enhancement was overestimated by neglecting the additional ionic-liquid capacitance that was induced by the back-gate voltage. In addition, capacitive coupling between the back and top gates through the large top-gate bonding pad may also lead to a significant underestimate of the back-gate capacitance in conventional dual-gated FET devices,[24] thus causing a nominal overestimate of the mobility. Unlike in conventional top-gated (or dual-gated) devices, where the top-gate electrode is directly on top of the channel, inducing capacitive coupling, the metal gate electrode used in the ionic-liquid gate, shown in Fig. 1a, is located within the plane of the device and separated by several tens of micrometers, causing no change in the capacitance. As a result, the ionic-liquid gate electrode in our devices is capacitively decoupled from the transport channel and the back-gate in the temperature range between 77 and 180 K, ruling out the possibility of any stray capacitance arising from the gate electrode that may inadvertently cause a nominal increase in the back-gate capacitance. Our $I_{ds}$-$V_{bg}$ measurements with a floating and grounded ionic-liquid gate, shown in Fig. 5a, yield identical results, indicating that the ionic-liquid gate electrode has no effect on the back-gate capacitance when the ionic liquid is frozen.

Figures 5(b-c) show the temperature dependence of the low-field field-effect mobility extracted from the $I_{ds}$-$V_{bg}$ curves in Fig. 5a. These data are compared to the expected mobility in the same device with no ionic liquid present using the expression $\mu = L/W \times dI_{ds}/dV_{bg}/(C_{bg}V_{ds})$. For this estimate, we used $L$ = 4.5 µm, $W$ = 0.7 µm, and $C_{bg}$ = 1.2 × 10$^{-8}$ Fcm$^{-2}$ for the back-gate capacitance. The field-effect mobility with no ionic liquid present was observed to decrease from 5 cm$^2$V$^{-1}$s$^{-1}$ to 0.3 cm$^2$V$^{-1}$s$^{-1}$ as the temperature decreased from 295 K to 140 K, following a simple activation temperature dependence depicted in Fig. 5c. In this case, the mobility decreases much more rapidly than if it were limited by scattering from charged impurities.[43] This suggests that the charge transport behavior is largely limited by the Schottky barriers at the contacts and does not reflect the intrinsic behavior of the carrier mobility. We can extract an effective Schottky barrier height of $\Phi$ ~ 66 meV from the temperature dependence of the extrinsic mobility $\mu \sim \exp(-\Phi/k_BT)$ in Fig. 5c. This is lower than published theoretical estimates for an ideal interface,[25] possibly due to band bending induced by an applied gate voltage. In sharp contrast to this behavior, the mobility in presence of the ionic-liquid gate increases from ~100 cm$^2$V$^{-1}$s$^{-1}$ to ~ 220 cm$^2$V$^{-1}$s$^{-1}$ as the temperature decreases from 180 K to 77 K at a carrier concentration between 7 × 10$^{12}$ and < 9 × 10$^{12}$ cm$^{-2}$ [determined from $n_{2D} = C_{bg}(V_{bg} - V_{th})$], following a $\mu \sim T^{-\gamma}$ dependence with $\gamma \approx 1$. We conclude that in this case, the mobility is limited by the intrinsic behavior of the channel. Of course, dielectric screening in the ionic-liquid gate could nominally increase the capacitive coupling.[43] Still, the observed qualitative change from thermally activated to "metallic" behavior caused by an ionic liquid gate, which acts as a top dielectric layer below the melting temperature, cannot be simply attributed to a possible

underestimation of the back-gate capacitance. Extrapolating the $\mu \sim T^\gamma$ fit with $\gamma \approx 1$ to the temperature $T=250$ K yields a mobility value of $\mu \approx 70$ cm$^2$V$^{-1}$s$^{-1}$, in good agreement with the mobility measured in the ionic liquid gate configuration at 250 K. This provides further evidence that the mobility measured in the back gate configuration is unlikely an artifact due to the underestimation of back gate capacitance. Moreover, as discussed in more detail below, our four-terminal measurements on a similar MoS$_2$ device show that the presence of the frozen ionic liquid, which forms an additional dielectric layer on top of the devices, does not substantially change the capacitance of the back-gate measurements.

Low-field field-effect channel mobility in this temperature range is affected by various scattering mechanisms, including scattering by acoustic phonons, optical phonons, as well as long range and short range disorder that is present both in the bulk and near the surfaces of the channel. Kaasbjerg *et al*. showed theoretically that the mobility due to acoustic and optical phonon scattering in monolayer MoS$_2$ increases with decreasing temperature following a $\mu \sim T^\gamma$ dependence, where the exponent $\gamma$ depends on the dominant scattering mechanism.[44] At relatively low temperatures ( < 100 K) , acoustic phonon scattering dominates, resulting in $\gamma = 1$. At higher temperatures, optical phonon scattering starts to dominate, and the exponent $\gamma > 1$ should cause a stronger temperature dependence of the mobility. On the other hand, the disorder-limited mobility decreases with decreasing temperature.[44-45] In our few-layer devices, the observed exponent ($\gamma \approx 1$) in the expression $\mu \sim T^\gamma$ for the temperature dependence of the mobility coincides with that of transport dominated by acoustic-phonon scattering. In this case, however, the mobility values are substantially lower than what would be expected from acoustic-phonon-limited mobility, and the temperature was high enough to excite not only acoustic, but also optical phonons. This behavior can be understood in a likely scenario, where the top ionic-liquid dielectric quenches phonon modes and thus reduces the γ value.

In order to verify that the extrinsic mobility of our ionic liquid gated MoS$_2$ FETs approaches the true channel mobility, we also performed four-terminal measurements. An AFM image of the four-terminal device is shown in the inset of Fig. 6a. We measured the conductance $G = I_{ds}/V_{inner}$, where $V_{inner}$ is the potential difference across the voltage probes, at a fixed drain-source bias while sweeping the back-gate voltage. $V_{inner}$ is kept below 70 mV during the measurement. Fig. 6a shows the conductance *versus* back-gate voltage of an 8 nm thick MoS$_2$ flake measured at various temperatures, with a back-gate voltage of up to $V_{bg}$ = 70 V. Under these conditions, we expect a charge carrier concentration of $n_{2D} = C_{bg}(V_{bg} - V_{th}) \sim 6.8 \times 10^{12}$ cm$^{-2}$ at $V_{bg}$ = 70 V, where the threshold voltage $V_{th} \approx$ -20 V. The field-effect mobility can be extracted from the $G$ vs. $V_{bg}$ curves in the 60 < $V_{bg}$ < 70 V range using the expression μ = $L_{inner}/W \times (1/C_{bg}) \times dG/dV_{bg}$, where $L$ = 3.0 μm is the distance between the two voltage probes, $W$ = 5.0 μm, and $C_{bg} = 1.2 \times 10^{-8}$ Fcm$^{-2}$ is the back-gate capacitance. The steps in the conductance may be caused by the back-gate tuning of the voltage contacts. The voltage electrodes are very wide in our geometry, and their separation is only about twice their width. In this case, the effective distance between them may change with applied gate-voltage. Fig. 6b shows the temperature dependence of the true channel mobility with and without the frozen ionic liquid as a dielectric capping layer. Before the ionic liquid was added, we observed a mobility increase from ≈ 40 cm$^2$V$^{-1}$s$^{-1}$ at 300 K to ≈ 390 cm$^2$V$^{-1}$s$^{-1}$ at 77 K following a $\mu \sim T^{-\gamma}$ dependence with γ≈1.7 in the entire temperature range, in good agreement with theoretical predictions for an MoS$_2$ monolayer.[44] Although our 8 nm thick sample is much thicker than a monolayer, back-gating has the strongest effect on the bottom MoS$_2$ layers. Moreover, charge screening also reduces the number of charge carriers in the top layers, especially at high back-gate voltages.[27, 46] The relatively low overall mobility values in this device may be attributed to additional extrinsic scattering mechanisms such as impurity scattering and scattering off surface polar optical phonons of the SiO$_2$ gate dielectric.[44] Interestingly, the overall mobility of the device with the ionic liquid is similar to that without the ionic liquid in the temperature range between 77 and 180 K, indicating that the ionic liquid as a capping top dielectric does not substantially impact the back gate capacitive coupling when it is frozen. On the other hand, the temperature dependence of the mobility weakens upon adding the ionic liquid, following a $\mu \sim T^{-\gamma}$ dependence with $\gamma \approx 1.2$ for $77 < T < 180$ K. This weaker temperature dependence of the channel mobility $\mu$ in presence of ionic liquid dielectric capping is consistent with that of the extrinsic mobility shown in fig. 5b, which can be attributed to phonon mode quenching caused by the ionic-liquid dielectric.[43] Note that the extrinsic mobility values in fig. 5b, which include the contact resistance, are in good agreement with the true channel mobility values in fig. 6b. This is a further demonstration that ionic liquid gating effectively creates highly transparent electrical contacts and reveals the intrinsic channel-limited device properties of MoS$_2$ FETs. This finding shows that the previously reported low mobility of Si-back-gated MoS$_2$ FETs, ranging typically between 0.1 – 10 cm$^2$V$^{-1}$s$^{-1}$, may not represent the intrinsic channel property, but was rather limited by non-ideal contacts, as pointed out by Popov *et al*.[25] It also demonstrates that phonon-limited mobility can be achieved without substantially reducing the disorder near the channel/dielectric interface.

Since the carriers induced by the ionic-liquid gate are close to the top surface of the MoS$_2$ channel, the interface properties in these devices may be different from back-gated devices with no ionic liquid, where the carriers are closer to the bottom surface. To rule out the possibility that the drastic difference in mobility between MoS$_2$ devices with and without an ionic liquid gate arises from differences at the channel/dielectric interface, we measured another few-nanometer thick (5 nm or ~7 layers) back-gated MoS$_2$ with patterned *n*-doping. As shown in the inset of Fig. 7a, both the electrical contacts and a small section of the channel near each electrode are covered by a 50 nm thick layer of positive resist (hydrogen silsesquioxane, HSQ), defined by standard electron beam lithography at a dose of 200 µC/cm$^2$, while the large portion of the channel is not covered by HSQ. At low degrees of cross-linking obtained at low electron beam dosages, Si-H bonds in HSQ are readily broken and release hydrogen, increasing electron density in MoS$_2$ near the contacts.[47] Consequently, HSQ covered contact regions are *n*-doped, while the bare channel region remains undoped. Fig. 7a shows selected transfer curves of the back-gated device at temperatures between 77 and 295 K, from which the field-effect mobility can be extracted. As shown in Fig. 7b, the mobility increases from ~ 75 cm$^2$V$^{-1}$s$^{-1}$ to ~ 180 cm$^2$V$^{-1}$s$^{-1}$ as the temperature drops from 295 K to 140 K, suggesting that mobility in this temperature range is largely limited by phonon scattering. Below 140 K, the mobility starts decreasing with decreasing temperature, which can be attributed to the reduction of the thermally assisted tunneling current through a Schottky barrier. Since the main difference between this device and other back-gated, few-nanometer-thick MoS$_2$ FETs is the *n*-doping of the contacts, the observed high mobility at room temperature and its phonon-limited temperature dependence above 140 K can be attributed to the Schottky barrier thinning by the surface doping in the contact regions, which is albeit not as effective as ionic-liquid gating. The slightly lower mobility in the ionic-liquid gated device (Fig. 5b) than in the HSQ contact-doped device (Fig. 7b) above 140 K may be attributed to the presence of added charge impurities from the ionic liquid.[48] This finding unambiguously demonstrates that the observed mobility enhancement in ionic-liquid-gated MoS$_2$

FETs is not an interface effect and cannot be simply attributed to the reduction of charge scattering in the transport channel.

**CONCLUSIONS**

In conclusion, we report the fabrication of ionic-liquid-gated $MoS_2$-based field-effect transistors with significantly higher mobilities than reported in comparable back-gated devices. We attribute the observed mobility enhancement to the ionic liquid, which acts as an ultrathin dielectric that effectively reduces the Schottky barrier thickness at the $MoS_2$/metal contacts by strong band-bending. The substantially reduced contact resistance in ionic-liquid-gated bilayer and few-layer $MoS_2$ FETs results in an ambipolar behavior with high ON/OFF ratios, a near-ideal subthreshold swing, and significantly improved field-effect mobility. Remarkably, the mobility of a 3-nm-thick $MoS_2$ FET with ionic-liquid-gating was found to increase from ~ 100 $cm^2V^{-1}s^{-1}$ to ~ 220 $cm^2V^{-1}s^{-1}$ as the temperature decreased from 180 K to 77 K. This finding is in quantitative agreement with the true channel mobility measured by four-terminal measurement, suggesting that the mobility is predominantly limited by phonon-scattering. It is remarkable that the high mobility has not been degraded by the presence of both long range disorder (*e.g.* from charged impurities) and short range disorder (*e.g.* interface roughness scattering) at the $MoS_2/SiO_2$ and $MoS_2$/ionic-liquid interface. The effect of Schottky barrier thinning on the performance of $MoS_2$ FETs was further verified by patterned *n*-doping of the contact regions using HSQ. More detailed studies of $MoS_2$ FETs with HSQ doped contacts are underway. Our study of ionic-liquid-gated $MoS_2$ FETs clearly demonstrates that previously observed low mobility values in monolayer and few-layer $MoS_2$ devices were largely caused by the large contact resistance. We found that phonon-limited mobility can be recovered through Schottky barrier thinning even in the presence of non-ideal channel/dielectric interfaces. We also surmise that the reported drastic mobility improvement in high-κ dielectric-capped MoS$_2$ FETs may be partially attributed to the Schottky barrier thinning caused by the doping of the fixed charges in the thin dielectric layer in addition to dielectric screening.

**METHOD**

Atomically thin MoS$_2$ flakes were produced by repeated splitting of a bulk crystal using a mechanical cleavage method and subsequently transferred onto degenerately doped silicon substrates covered with a 290 nm-thick thermal oxide layer. An optical microscope was used to identify thin flakes, which were further characterized by Park-Systems XE-70 non-contact mode atomic force microscopy (AFM). MoS$_2$ IL-FET devices were fabricated by first patterning the source, drain and gate electrodes, consisting of 5 nm of Ti covered by 50 nm of Au, using standard electron beam lithography and electron beam deposition. A small droplet of the DEME-TFSI ionic liquid (Sigma Aldrich 727679) was then carefully applied onto the devices using a micromanipulator under an optical microscope, covering the MoS$_2$ layer and the source, drain and gate electrodes. The ionic liquid gate forms by self-assembly. Electrical properties of the devices were measured by a Keithley 4200 semiconductor parameter analyzer in a Lakeshore Cryogenic probe station after dehydrating the ionic liquid under high vacuum (~1×10$^{-6}$ Torr) for 48 hours.

Acknowledgements

This work was supported by NSF (No. ECCS-1128297). DT was supported by the National Science Foundation Cooperative Agreement No. EEC-0832785, titled ''NSEC: Center for High-rate Nanomanufacturing.'' Part of this research was conducted at the Center for Nanophase Materials Sciences under project # CNMS2011-066. The authors also thank Michael Fuhrer and Kristen Kaasbjerg for helpful discussion.

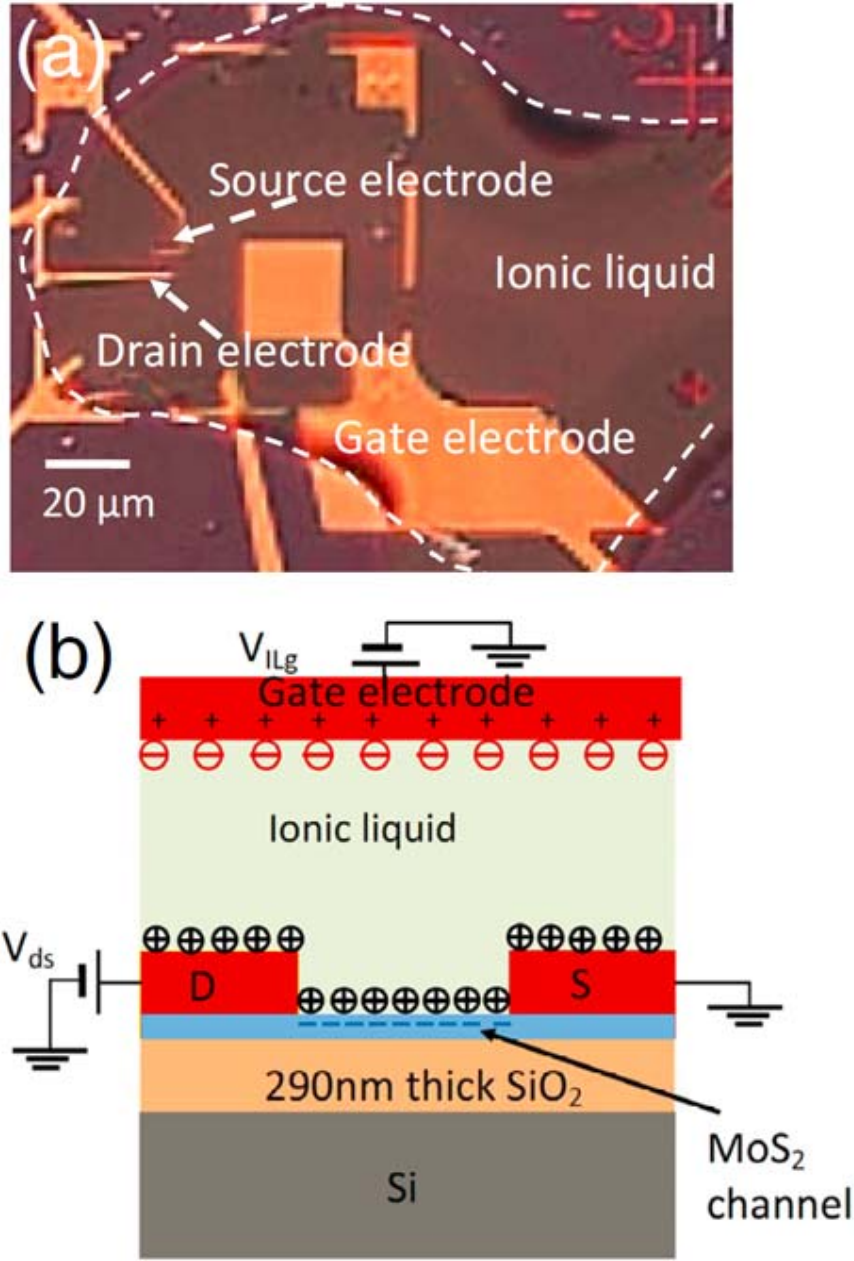

**Figure 1.** (a) Optical micrograph of a typical ionic-liquid-gated $MoS_2$ FET. The contour of the ionic liquid drop covering the $MoS_2$ channel and the in-plane gate-electrode are marked by white dotted lines. The scale bar is 20 μm. (b) Schematic illustration of the working principle of an ionic-liquid-gated $MoS_2$ FET.

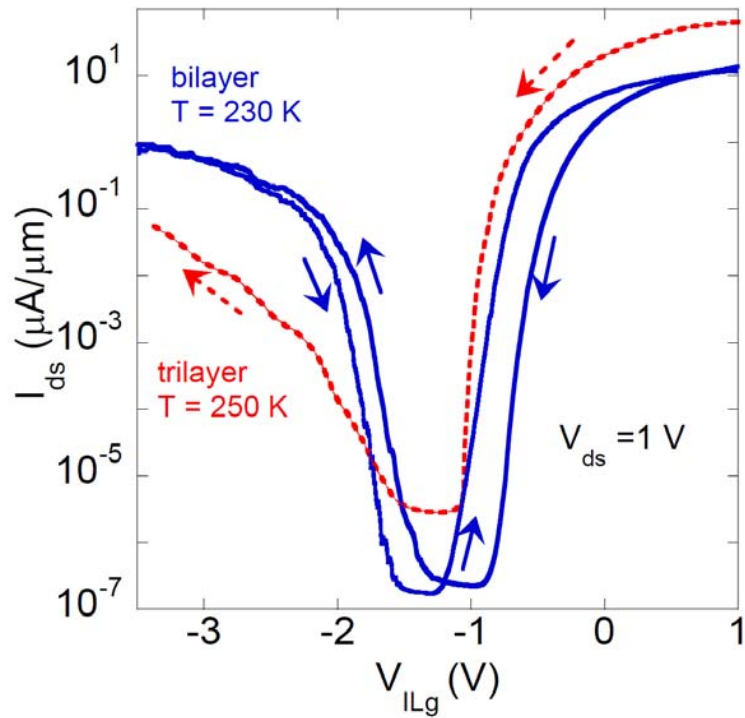

**Figure 2.** Transfer characteristics of representative bilayer and trilayer $MoS_2$ ionic-liquid-gated FETs measured at the drain-source bias $V_{ds}$ = 1 V.

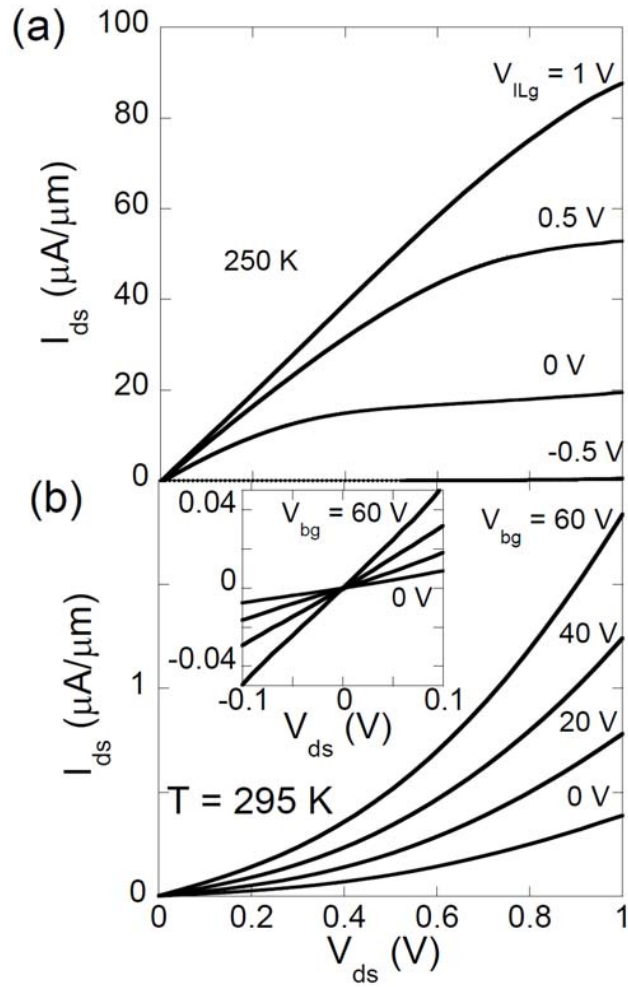

**Figure 3.** Comparison of the output characteristics of the trilayer MoS$_2$ device used in Fig. 2, measured in the ionic-liquid-gate and back-gate (without ionic liquid) configurations. (a) Drain-source current $I_{ds}$ as a function of the drain-source bias $V_{ds}$ at ionic-liquid-gate voltages between -0.5 and 1 V. (b) $I_{ds}$ as a function of $V_{ds}$ at selected back-gate voltages between 0 and 60 V before the ionic liquid was deposited. The inset in (b) shows the magnified low-bias region in this panel.

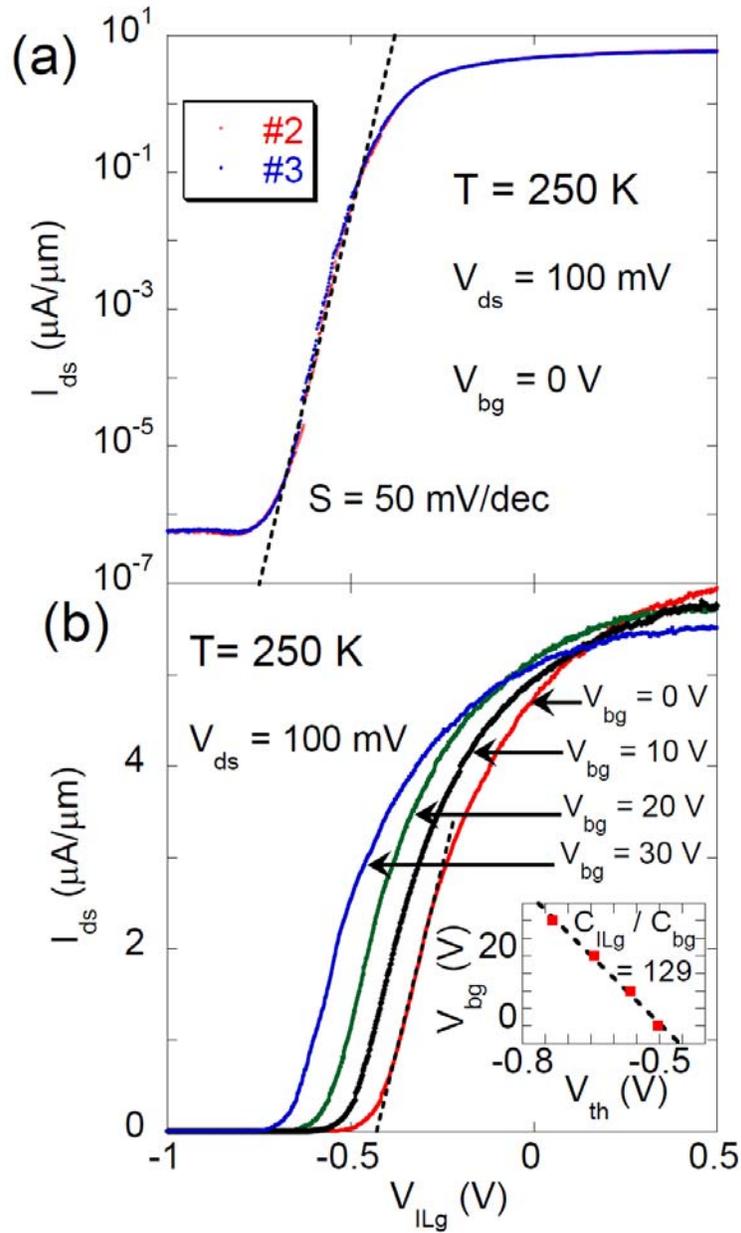

**Figure 4.** (a) Transfer characteristics of the identical trilayer $MoS_2$ device in two separate runs, where the ionic-liquid-gate voltage was swept at $V_{ds}$ = 100 mV, $V_{bg}$ = 0 V and $T$ = 250 K. (b) Transfer curves of the identical ionic-liquid-gate device measured at various back-gate voltages between 0 and 30 V. The inset in (b) shows the back-gate voltage *vs.* the threshold voltage of the transfer curves.

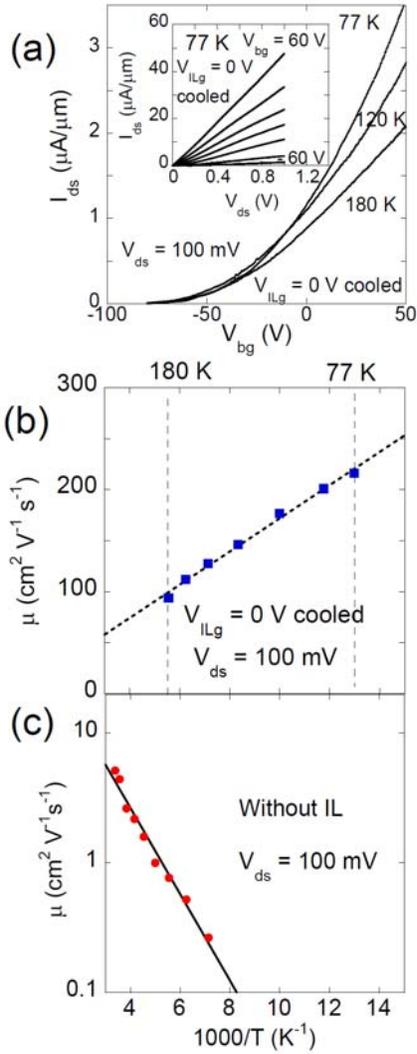

**Figure 5.** (a) Transfer curves of a 3.3 nm thick (5 layer) MoS$_2$ FET measured in the back-gate configuration with the drain-source bias $V_{ds}$ = 100 mV and the ionic-liquid-gate voltage kept at 0 V. The observations in the temperature range between 77 and 180 K were performed after the device had been cooled down from 250 to 77 K. The inset in (a) shows the output characteristics of the device measured at back-gate voltages between -60 and 60 V at $T$ = 77 K. (b) Temperature dependence of the field-effect mobility extracted from the transfer characteristics in (a). (c) Temperature dependence of the field-effect mobility of the same device before the ionic liquid was added.

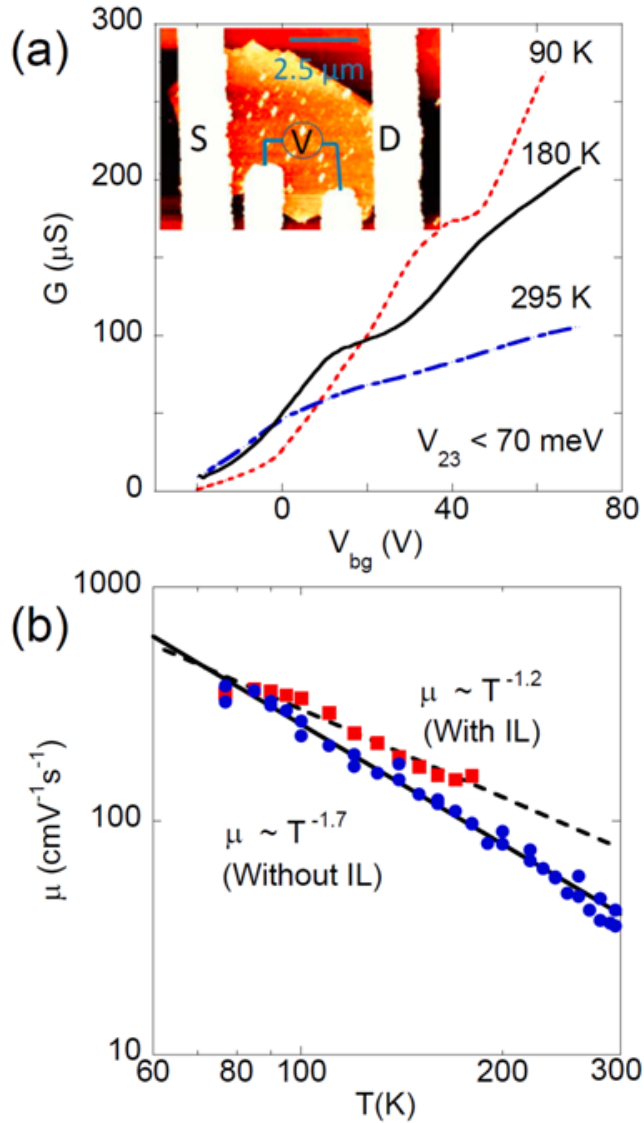

**Figure 6.** Four-terminal electron transport in a back-gated 8 nm thick MoS$_2$ FET with and without ionic liquid. (a) Conductance as a function of back-gate voltage measured at different temperatures with no ionic liquid present. The inset shows an AFM image of the four-terminal device. (b) Temperature dependence of the true channel mobility derived from the four-terminal measurements in presence and absence of the ionic liquid.

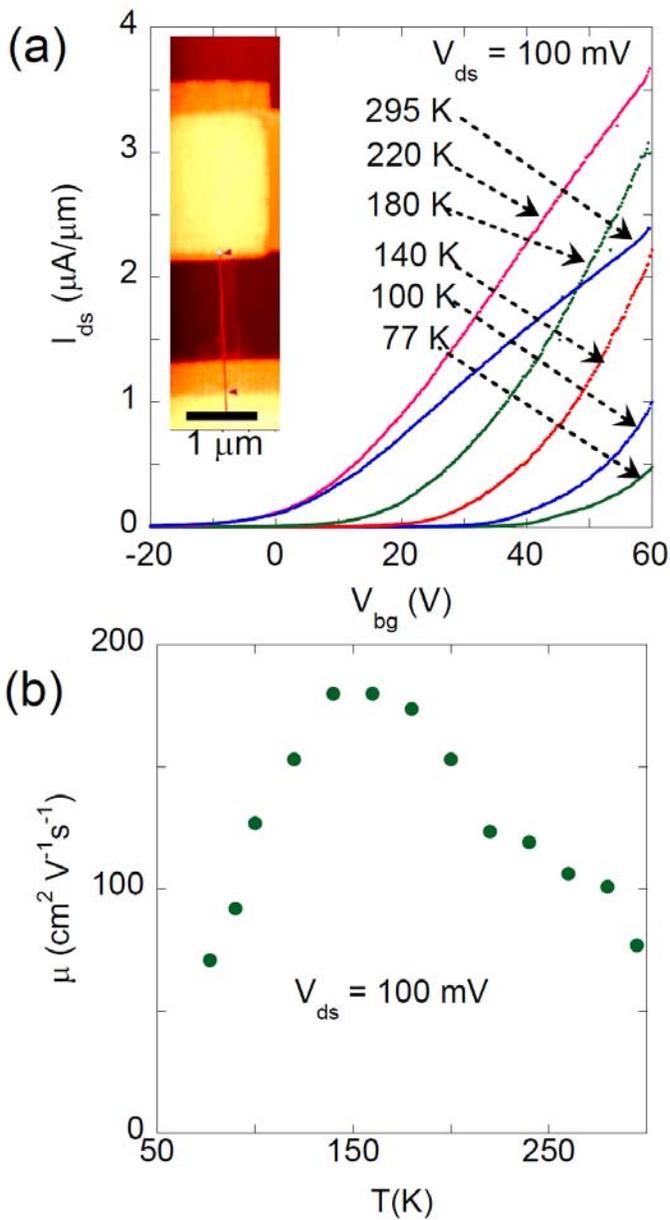

**Figure 7**. (a) Transfer characteristics of a back-gated 5 nm thick $MoS_2$ FET measured at the drain-source bias $V_{ds}$ = 100 mV, for various temperatures. The inset of (a) shows an AFM image of the device with its electrical contacts covered by HSQ while large portion of the channel is bare. (b) Field-effect mobility of the device as a function of temperature.